\documentclass{aastex631}

\usepackage{multirow}
\usepackage{threeparttable}
\begin{document}

\title{Detection of a new GeV source in the outer region of the Coma cluster: a signature of external accretion shock ?}

\author[0000-0003-4907-6666]{Xiao-Bin Chen}
\affiliation{School of Astronomy and Space Science, Nanjing University, Nanjing 210023, China; xywang@nju.edu.cn; xbc@smail.nju.edu.cn}
\affiliation{Key laboratory of Modern Astronomy and Astrophysics (Nanjing University), Ministry of Education, Nanjing 210023, China}

\author[0000-0002-5582-8265]{Kai Wang}
\affiliation{School of Astronomy and Space Science, Nanjing University, Nanjing 210023, China}
\affiliation{Key laboratory of Modern Astronomy and Astrophysics (Nanjing University), Ministry of Education, Nanjing 210023, China}

\author[0000-0002-6036-985X]{yi-Yun Huang}
\affiliation{School of Astronomy and Space Science, Nanjing University, Nanjing 210023, China}
\affiliation{Key laboratory of Modern Astronomy and Astrophysics (Nanjing University), Ministry of Education, Nanjing 210023, China}

\author[0000-0001-6863-5369]{Hai-Ming Zhang}
\affiliation{Guangxi Key Laboratory for Relativistic Astrophysics, School of Physical Science and Technology, Guangxi University, Nanning 530004, China}

\author{Shao-Qiang Xi}
\affiliation{Key Laboratory of Particle Astrophysics \& Experimental Physics Division \& Computing Center, Institute of High Energy Physics, Chinese Academy of Sciences, 100049 Beijing, China}

\author[0000-0003-1576-0961]{Ruo-Yu Liu}
\affiliation{School of Astronomy and Space Science, Nanjing University, Nanjing 210023, China}
\affiliation{Key laboratory of Modern Astronomy and Astrophysics (Nanjing University), Ministry of Education, Nanjing 210023, China}

\author[0000-0002-5881-335X]{Xiang-Yu Wang}
\affiliation{School of Astronomy and Space Science, Nanjing University, Nanjing 210023, China}
\affiliation{Key laboratory of Modern Astronomy and Astrophysics (Nanjing University), Ministry of Education, Nanjing 210023, China}

\begin{abstract}
The supersonic flow motions
associated with infall of baryonic gas toward sheets and
filaments, as well as cluster mergers, produces large-scale shock waves. The shocks associated with galaxy clusters can be classified mainly into
two categories: internal shocks appear in the hot
intracluster medium within the viral radius, and  external accretion shocks
form in the outer cold region well outside of the virial radius. Cosmic-ray (CR) electrons and/or protons accelerated by these shocks are expected to produce gamma-rays through inverse-Compton scatterings (ICS) or inelastic $pp$ collisions respectively. Recent studies have found a spatially extended GeV source within the virial radius, consistent with the internal shock origin.  
Here we report the detection of a new GeV source at a distance of about 2.8$^\circ$ from the center of the Coma cluster through the analysis of 16.2 years of Fermi-LAT data.  The hard spectrum of the source, in agreement with the ICS origin, and its location in a large-scale filament of galaxies points to the external accretion shock origin. The gamma-ray ($0.1-10^3$ GeV) luminosity of the source, $1.4\times 10^{42}~ {\rm erg~s^{-1}}$, suggests that a fraction  $\sim 10^{-3}$ of the kinetic energy flux through the shock-surface is transferred to  relativistic CR electrons.
\end{abstract}


\section{Introduction} \label{sec:intro}

Galaxy clusters, the largest gravitationally bound structures in the Universe, are thought to form through mergers
and accretion of smaller structures. The supersonic flow motions associated with cluster mergers and accretion naturally induce
shocks, which have been extensively studied using cosmological hydrodynamic simulations \citep{Miniati2000,Ryu2003,Pfrommer2006,Ha2023}. The shocks can be classified mainly
into two categories  \citep{Ryu2003}. One is internal shocks that appear
in the hot ($T\sim 10^8 {\rm K}$) intracluster medium (ICM) within the virial radius as the merging components approach each other
at slightly supersonic  speed.
Another is external accretion shocks that form in the outer region well outside of the virial radius, where the  the cold ($T\sim 10^4 {\rm K}$) gas in void regions and
the warm-hot ($T\sim 10^5-10^7 {\rm K}$) intergalactic medium in filaments accrete onto clusters. Accretion shocks  are typically
much stronger with highly supersonic  speed since they develop in cold
external cluster regions. 
As for typical astrophysical shocks, these cosmological shocks are collisionless, and hence are expected to
accelerate cosmic ray (CR) protons and electrons to very
high energies via diffusive shock acceleration\citep{LoebWaxman2000}. Interactions between CR particles and the ambient
medium, radiation, and magnetic fields may generate nonthermal emission in a wide range of wavelengths.

%Radio observations prove the existence of relativistic particles and magnetic field associated with the ICM through the presence of extended synchrotron
%emission in the form of radio halos and peripheral relics. The accelerated CRs can produce
%HE gamma-rays through leptonic and hadronic radiative processes.  Studying clusters of galaxies in the gamma-ray band is challenging
%but can provide important constraints on the efficiency of particle
%acceleration, magnetic confinement, and radiation processes of CRs
%(Blasi, Gabici \& Brunetti 2007). 

The Coma cluster of galaxies is the nearest massive
clusters at a distance of $\sim 100$ Mpc. It shows evidence of
efficient particle acceleration, as suggested by the presence
of a giant radio halo and radio relics \citep{Thierbach2003,Brown2011,vanWeeren2019SSRv}. 
Fermi Large Area Telescope (Fermi-LAT) collaboration found two low-significance structures within the viral
radius of the Coma cluster using 6
yr data \citep{Ackermann2016Coma}.  \cite{Xi2018Coma} reported the discovery of significant GeV gamma-ray emission from the Coma cluster with an unbinned likelihood analysis of the 9 years of Fermi-LAT  data.  For the first time, they also found tentative evidence that the gamma-ray emission  is spatially extended.  Later works with  more data and detailed analyses \citep{Adam2021Coma,Baghmanyan2022Coma} 
confirmed that a significant gamma-ray signal is observed roughly within the virial radius of  the Coma cluster and the  source is spatially extended. This gamma-ray source is located within the viral radius, consistent with the internal shock origin.

Nonthermal emission in the outer region of galaxy
clusters beyond the virial radius has long been postulated as direct evidence for CR acceleration at external
accretion shocks\citep{Keshet2003Shock,Ryu2003}. Theoretically, it is
expected that gamma-ray emission  is dominated by the hadronic process in the cluster core region,
whereas in the outskirts region the IC emission is the
dominant component {\citep{Miniati2001, Pinzke2010}}.  In particular, gamma-ray \citep{LoebWaxman2000,Totani2000,Scharf2002, Keshet2003Shock, Miniati2003NumModel,Pinzke2010} and hard X-ray  radiation \citep{Enblin1999HXR,Kushnir2010HXR} could
possibly be produced by the inverse-Compton (IC) scattering of cosmic microwave background (CMB) photons
off CR electrons accelerated at accretion shocks. Radio synchrotron radiation could also originate from accretion
shocks as well \citep{Bonafede2022LOFAR}.  Although
a few studies  found preliminary evidence for a large gamma-ray ring  around the Coma cluster from  the 
VERITAS mosaic and Fermi-LAT data \citep{Keshet2017, KeshetReiss2018}, more clear observational evidence would be required
to firmly support CR  acceleration at accretion shocks.
In this {\em Letter}, we report the detection of a gamma-ray source in the outer region of the Coma
cluster   at a statistic significance  of $\sim 5\sigma$ by using 16.2 years of Fermi-LAT data.  The spatial position and spectral/morphological properties of the source  suggest   accretion shock as the origin of the gamma-ray emission.

The rest part of the paper is organized as follows. We present a description of the data analysis and show the morphological and spectral results of the new gamma-ray source in Section 2. Then we discuss the accretion shock origin of the source in Section 3. Finally, we give conclusions and discussions in Section 4.
\section{Fermi-LAT data analysis}

\subsection{Data selection and background models}

We used 16.2 yr of Fermi-LAT data from August 2008 to October 2024 to study the GeV emission near Coma cluster. The event class {\tt\string P8R3\_ULTRACLEANVETO} (evclass = 1024) and event type FRONT + BACK (evtpye = 3) are used. This is the cleanest Pass 8 event class and recommended to check for extra-galactic diffuse analysis \footnote{\url{https://fermi.gsfc.nasa.gov/ssc/data/analysis/documentation/Cicerone/Cicerone_Data_Exploration/Data_preparation.html}}.
To account for the diffuse emission, we modelled the
Galactic diffuse emission model ({\tt\string gll\_iem\_v07.fits}) with isotropic component({\tt\string iso\_P8R3\_ULTRACLEANVETO\_V3\_v1.txt}) relevant to
the {\tt\string ULTRACLEANVETO} event class. We use recommended time selection of $\rm (DATA\_QUAL>0)\&\&(LAT\_CONFIG == 1)$.
To minimize the contamination from the Earth limb, the maximum zenith angle is set to be $90^{\circ}$. 

The data selection was within a region of interest (ROI) of $20^\circ$ around the center of the Coma cluster at (R.A., Dec.) = (194.95$^\circ$, 27.98$^\circ$) with an energy range between 100 MeV and 1 TeV. We include the Galactic diffuse emission(GDE), isotropic emission and all sources listed in the fourth Fermi-LAT catalog \citep{Fermi4FGL-DR4_2023} in the background model. 
All sources within $6^{\circ}$ of the center, as well as the GDE and isotropic emission components, are left free.
The maximum likelihood test statistic (TS) is used to estimate the significance of gamma-ray sources, which is defined by $\rm TS = 2(ln\mathcal{L}_1-ln\mathcal{L}_0)$, where $\rm ln\mathcal{L}_1$ and $\rm ln\mathcal{L}_0$ are maximum likelihood values for the background with target source and without the target source (null hypothesis). 
In this work, the publicly available software \textit{Fermitools} (v2.2.0) and the \textit{Fermipy} tool (version 1.2.2) is used to perform the data analysis.

\subsection{Morphological and spectral analysis}

\begin{figure}
    \centering
    \includegraphics[width=0.32\linewidth]{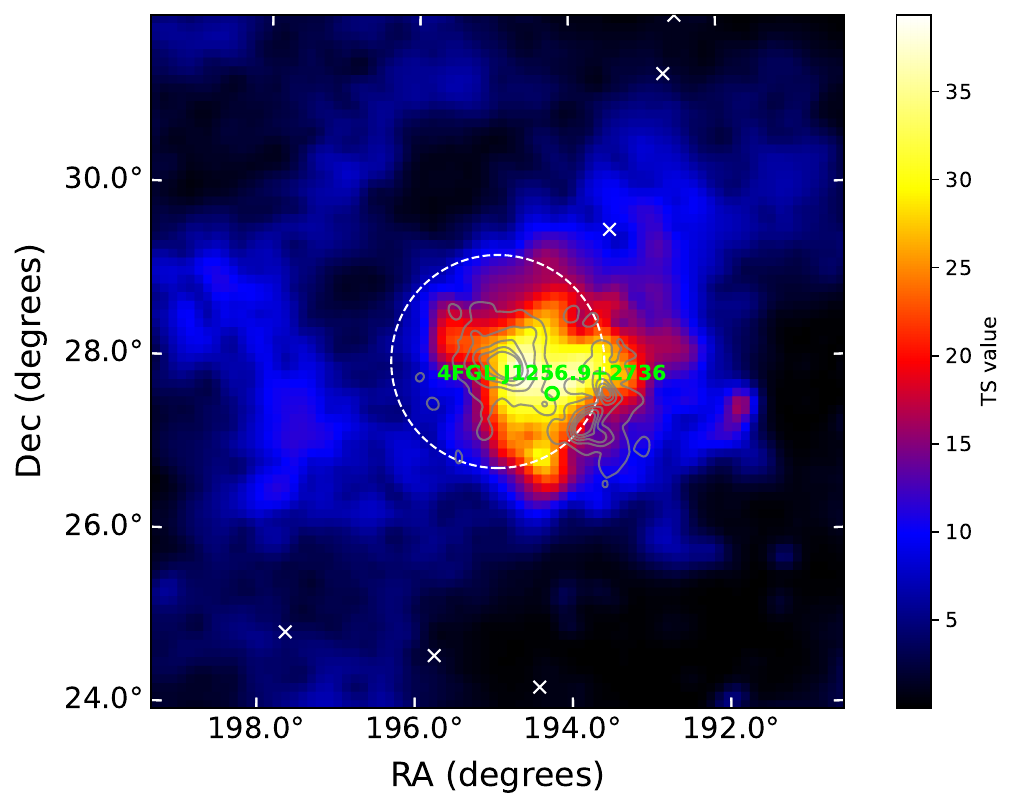}
    \includegraphics[width=0.32\linewidth]{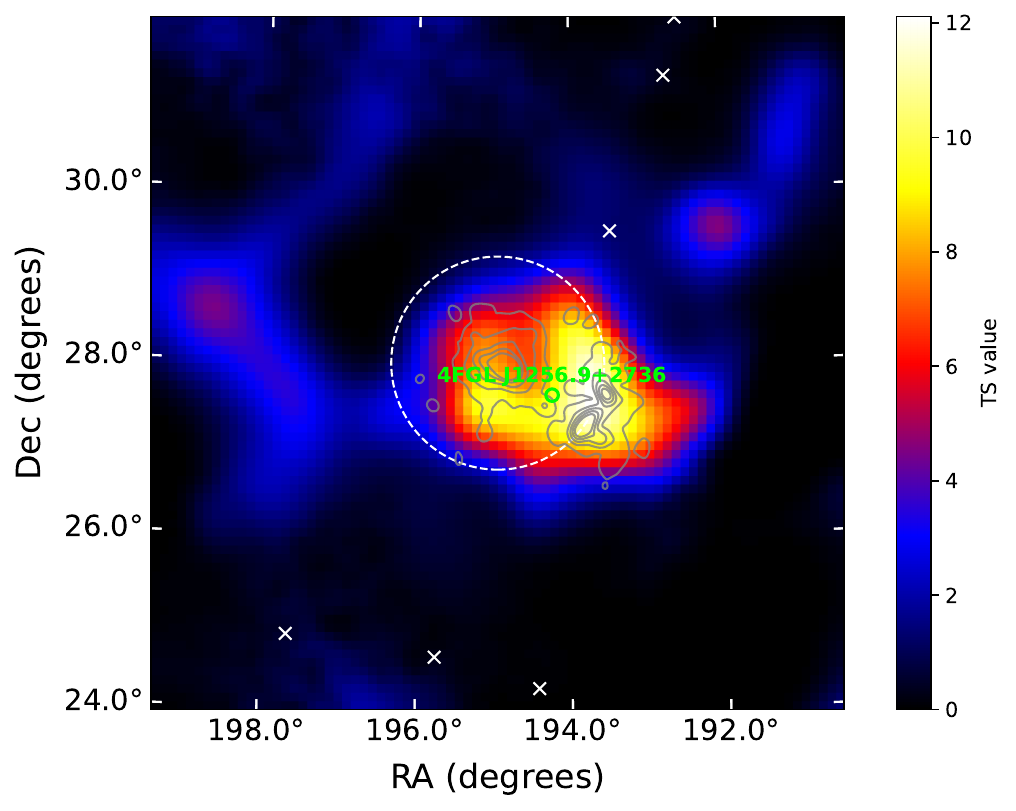}
    \includegraphics[width=0.32\linewidth]{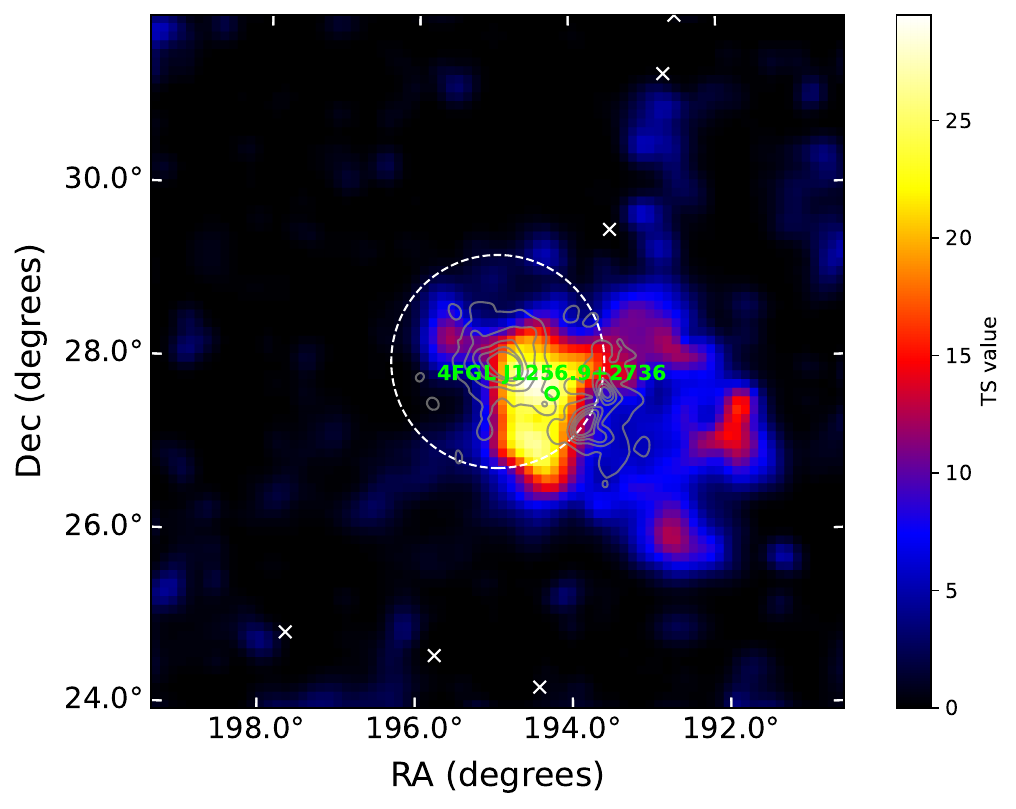}
    \caption{TS map of a $8^\circ \times 8^\circ$ region centered on the Coma cluster (left-hand panel: 100MeV-1TeV, middle panel: 100 MeV-500MeV, right-hand panel: 500MeV-1TeV). The dashed white circle represents the virial region of the Coma cluster ($\theta_{200}=1.23^\circ$). The gray contours correspond to measurements of the Coma cluster using WSRT at a central frequency of 352\,MHz \citep{Brown2011}. The green circle shows the location of  4FGL J1256.9+2736 and the white crosses show the location of other 4FGL sources.}
    \label{fig:TSmap3}
\end{figure}

% To study the outer region of the  Coma cluster, we choose a large size for the data analysis. 
% In the update  4FGL catalog (4FGL-DR4 \citep{Fermi4FGL-DR4_2023}),  a source in the direction of the Coma cluster, named 4FGL J1256.9+2736, is included. 
% To study the outer region of the Coma cluster, we first analyzed the data for 100 MeV–1 TeV energy band by removing 4FGL J1256.9+2736 from the source list as a baseline model.
Based on the 4FGL catalog, there is a 4FGL point-like source, named 4FGL J1256.9+2736, that is roughly coincident with Coma cluster in position. The TS map is generated by removing this source from the background model. The resulted TS maps of the $ 8^\circ \times 8^\circ$  region are shown in Fig. \ref{fig:TSmap3} for 100 MeV–1 TeV, 100 MeV–500MeV, and 500 MeV–1 TeV energy bands from left to right, respectively. In order to search for gamma-ray emission in the outer region of the Coma cluster, we have chosen a larger region of the Fermi-LAT data for the analysis  than previously considered \citep{Xi2018Coma,Adam2021Coma,Baghmanyan2022Coma}.

The morphology of the GeV emission is studied using the 4FGL J1256.9+2736 as the baseline model. We consider four spatial templates: 
(1) A point-like source (PS), i.e., 4FGL J1256.9+2736. We use the \textit{Fermipy} tool to quantitatively reevaluate the location of this source. 
(2) The radio template. We consider a template  assuming that the gamma-ray flux distribution traces the observed radio radio emission, since the radio emission traces relativistic electrons via synchrotron emission and could be associated with the gamma-ray signal. The radio emission template is based on the measurements of the Coma radio halo and relics using the Westerbork Synthesis Telescope (WSRT) at 352 MHz by \cite{Brown2011}. 
(3) A disk template. The best-fitted center of the disk is located at (R.A., Dec.) = ($194.25^\circ \pm 0.15^\circ$, $27.56^\circ \pm 0.17^\circ$) and the best-fitted extension is $R_{68} = 0.89^{\circ} \pm 0.11^{\circ}$. Here, we use the \textit{extension} tool in \textit{Fermipy} to measure the position and size of the extended source \footnote{We avoided self-inconsistent results caused by program settings by customizing parameters, see as \url{https://github.com/fermiPy/fermipy/issues/555}}. This tool computes TS of the extension hypothesis defined as $\rm TS_{ext} = 2 \times (\log \mathcal{L}_{ext} - \log \mathcal{L}_{PS})$, where $\rm \log \mathcal{L}_{ext}$ and $\rm \log \mathcal{L}_{PS}$ are likelihood values of the extended and point-like source models, respectively.
We found that the extension TS value of the disk model is $\rm TS_{ext}=28.046$. 
(4) A Gauss template. The best-fitted center of the disk is located at (R.A., Dec.) = ($ 194.29^\circ \pm 0.16^\circ$, $27.73^\circ \pm 0.16^\circ$) and the best-fitted extension is $R_{68} =  0.89^{\circ} \pm 0.19^{\circ}$ with $\rm TS_{ext}=26.100$.
Compared with the point-like source model, all of the disk, Gauss model and radio emission model improve the fitting significantly (see Table \ref{tab:label} for more details), which is consistent with previous analysis \citep{Xi2018Coma,Adam2021Coma,Baghmanyan2022Coma}. 
Since the disk template has the best improvement on likelihood, our subsequent analysis will be based on this template.

\begin{figure}
    \centering
    \includegraphics[width=0.9\linewidth]{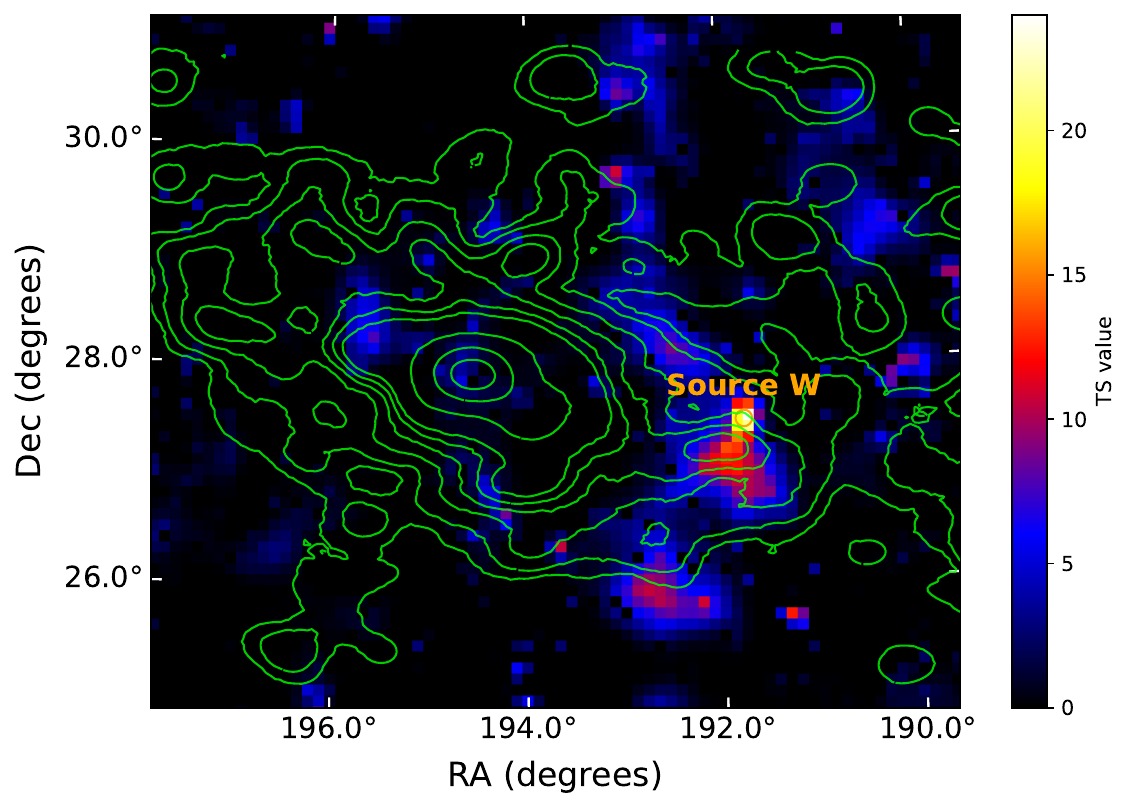}
    \caption{The residual map in 500 MeV-1 TeV of the Coma cluster region after subtracting the disk component in the core region. The green contours show the SDSS galaxy density distribution. }
    \label{fig:excessmap}
\end{figure}

Besides the gamma-ray emission presented above, which is within the viral radius $\theta_{200}$ (here the
subscript 200 refers to an enclosed density 200 times above the critical density of the Universe) of the Coma cluster, we found that there is a significant excess to the west of this extended source, about 2.8$^\circ$ from the Coma cluster, particularly evident in the 500 MeV–1 TeV map. In Figure \ref{fig:excessmap}, we present the residual map after subtracting the disk component from Coma cluster in the 500 MeV–1 TeV range, showing the presence of residual emission. 
%The residual emission has an arc-like shape with a brighest point. 
Therefore, we tested a two-component model, adding a point-like source to the original disk model. The best-fitted position of the new point-like source is (R.A., Dec.)$ = (191.772^{\circ} \pm 0.035^{\circ}, 27.465^{\circ} \pm 0.045^{\circ})$. We find that the TS value is increased by $\Delta {\rm TS\simeq 26}$ for the disk plus a point-like source model compared to the single disk model (see Table \ref{tab:label}), supporting
the presence of an additional gamma-ray source at a statistic significance of about $5\sigma$ (hereafter, we name the point-like source as the source "W"). To assess the robustness of this additional source, we investigate how the detection significance of the
source W changes in the systematic checks. The TS value of the source ranges from 23.5 to 28.4 (see Appendix A for more details). To test the extension of the source W, we consider a uniform disk  template to fit the source, but find that there is no improvement relative to the point-like source model. This could be due to the low significance of the source and the low
Fermi-LAT angular resolution. 

We examine whether there is any known counterpart at the position of the source W.  
Since most of the high-latitude gamma-ray background sources are  active galactic nuclei (AGNs) and they are sources of
radio emission, we  search for possible radio sources
associated with the source W. We find no bright radio sources within 99\% containment radius of the source W (see Appendix B for more details). 

We compare the TS map of the gamma-ray source to the galaxy density distribution, as shown in Fig. \ref{fig:excessmap}.
We select galaxies with spectroscopic information from the SDSS database \footnote{\url{https://skyserver.sdss.org/dr18}} and use this catalog to construct a galaxy density map around the Coma region \citep{Adam2021Coma}.
The source W is located in a galaxy filament in the west  of   the Coma cluster and is spatially coincident with a local overdensity  of the galaxy distribution.  Thus this source could be connected to the Coma cluster and its large-scale environment. 
%This agrees with the picture that gas in void regions accretes onto the filaments of galaxies and produces an accretion shock in the outer region of the galaxy cluster.

\begin{figure}
    \centering
    \includegraphics[width=0.9\linewidth]{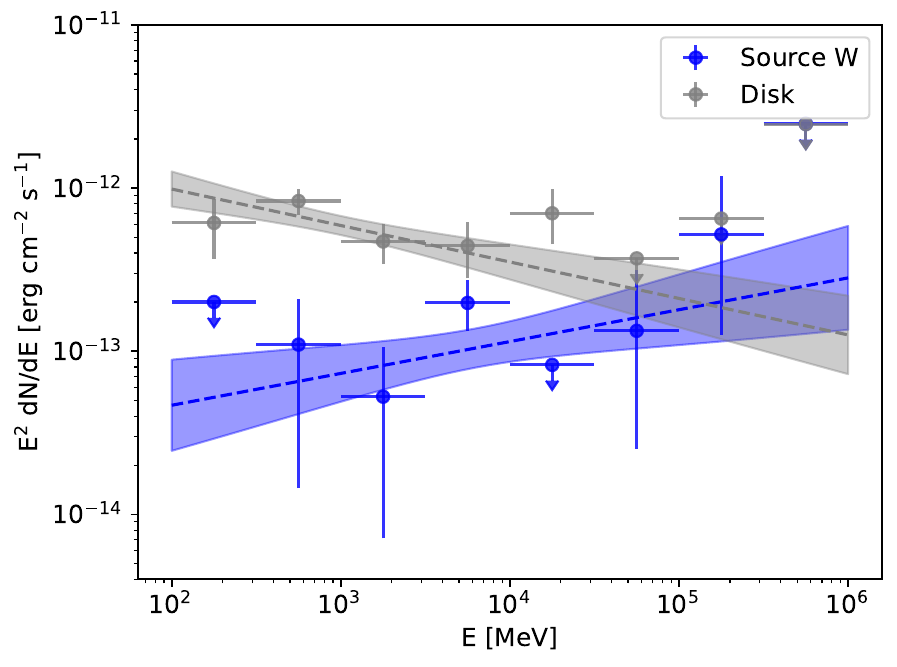}
    \caption{The spectral energy distribution (SED) of the disk component  (in gray) and the point-like source component (in blue).  The shaded areas mark the 1$\sigma$ errors of the fitted spectral model. }
    \label{fig:point_sed}
\end{figure}

Considering the disk plus a point-like source template, we calculate the spectral energy distribution
(SED) of the two components, which are shown in Fig. \ref{fig:point_sed}. Fitting with a power-law function, the photon index is $\alpha=-1.81\pm0.21$ for the point-like source component. The spectrum is quite hard compared with the gamma-ray emission from the disk component, whose photon index is $-2.22 \pm 0.11$. The hard spectrum motivate us to check the highest energy photons from the point-like source and find that two photons have energy of about 40 GeV and 110 GeV respectively. We use the {\tt\string gtsrcprob} tool to  obtain the probability of each photons coming from the point-like source  in our model, which is shown in Table \ref{tab:gtsrcprob}. The high probability of the association proves that the highest-energy photons belong to the point-like source and explain its hard spectrum. The energy flux of the disk component and the point-like source in 0.1–1000 GeV are respectively  $(3.85 \pm 0.65)\times 10^{-12}~{\rm erg~cm^{-2}~s^{-1}}$ and $(1.20 \pm 0.58)\times 10^{-12}~{\rm erg~cm^{-2}~s^{-1}}$, leading to a total gamma-ray
luminosity of $L_\gamma=(4.61 \pm 0.77)\times 10^{42}~{\rm erg~s^{-1}}$ and $L_\gamma=(1.44 \pm 0.69)\times 10^{42}~{\rm erg~s^{-1}}$ for the disk and point-like source, respectively. 

We examine the flux variability of the  source W by computing the light
curves in four and eight time bins over 16.2 yr, for events in the
energy range 0.1–1000 GeV. The result is shown in the Appendix C. We then use a likelihood-based
statistic to test the significance of the variability \citep{Nolan2012ApJS}. No evidence of
the flux variability is found (see Appendix B for more details).

\begin{table}
    \centering
    \begin{threeparttable}
    \begin{tabular}{c|ccc}
    \hline
      Spatial model  & ~ Spectral index ~& TS & $\Delta$TS   \\
    \hline
      PS model      & 2.69 $\pm$ 0.17 &29.11 & -     \\
      \hline
      Radio model            & 2.36 $\pm$ 0.13 &47.26 &    15.56   \\
      \hline
      Disk                   & 2.22 $\pm$ 0.10 &57.98 &     26.22  \\
      \hline
      Gauss                  &   2.21 $\pm$ 0.10      &   57.10   &     25.31  \\
      \hline
      Radio model            & 2.35 $\pm$ 0.13 & 47.19 &    \multirow{2}{*}{41.10}\\
      Source W & 1.81 $\pm$ 0.20 &25.66 &      \\
      \hline
      Disk                   & 2.22 $\pm$ 0.11 & 59.98 &  \multirow{2}{*}{  52.02}\\ 
      Source W & 1.81 $\pm$ 0.20 &25.48 &      \\
      \hline
      Gauss                   & 2.22 $\pm$ 0.10 & 60.54 &  \multirow{2}{*}{ 47.45}\\ 
      Source W & 1.78 $\pm$ 0.21 &24.61 &      \\
      \hline
    \end{tabular}
    \tablecomments{
    The $\Delta$TS represents the improvement in the test statistic when comparing the fit of PS model (the first line) to alternative models. 
    }
    \end{threeparttable}
    \caption{The result of the morphological analysis of the Coma cluster region in 100MeV-1TeV.}
    \label{tab:label}
\end{table}

\begin{table*}
    \centering
    \begin{tabular}{c|cccc}
    \hline
      Photon energy (MeV)   & Source W  &  Galactic diffuse emission  & Isotropic emission  &  Other sources\\
    \hline
      111456   & 96.5\% & 1.0\% & 2.4\% & 0.1\%\\
      43886   & 99.3\% & 0.2\% & 0.5\% &  0.0\%\\
      % 6504   & 0.897 & 0.036 & 0.066 &   0.001\\
      % 3969   & 0.826 & 0.070 & 0.103 & 0.001\\
    \hline
    \end{tabular}
    \caption{The probability of association with various sources for the two highest energy photons from the direction of the source W.}
    \label{tab:gtsrcprob}
\end{table*}

\section{Accretion shock origin for the source W}

The hard spectrum of the source W and its spatial coincidence with a local overdensity  of the galaxy distribution  suggest that it could result from an accretion shock. The accretion shock is characterized by high sonic velocity  with a Mach number $M_s\gg 1$,  and is expected to be efficient accelerators of cosmic ray protons and electrons.   Due to a low gas density  in the cluster outskirts ($n\sim 10^{-4}{\rm cm^{-3}}$), the hadronic emission from the cosmic-ray protons is expected to be low, so the gamma-ray spectrum is dominated by the IC emission of primary CR electrons accelerated by  accretion shocks. The energy distributions of CRs is described by a power-law function, $dN(E)/dE\propto E^{-p}$,  with $p$ determined
by the shock Mach number $M_s$, which approaches $p=2$ for $M_s \gg1$ \citep{Brunetti2014IJMPD}.  The cosmic-ray electrons can be accelerated  up to energies of tens of TeV, and the resulting IC emission by scattering  the CMB photons   extends to multi-TeV \citep{Keshet2003Shock}. 
These electrons lose a small fraction of energy into synchrotron emission since the magnetic field at the accretion shock is expected to be weak, $\le 0.1\mu {\rm G}$.  Because the cooling time of the GeV-emitting electrons (via the IC process), $t_{\rm cool}$  is short compared to the cluster
dynamical time, $t_{\rm dyn} \sim 1 {\rm Gyr}$, these electron lose  all their energy through IC emission during the dynamic time. Therefore, the resulting IC spectrum provides roughly equal energy flux per decade in photon energy for an electron spectrum with $p=2$ , i.e. $\nu F_\nu \propto \nu^0$ in the photon energy
range 100 MeV-1 TeV.  The measured spectral index ($\alpha=-1.81\pm0.21$) of the source W agrees with the predicted flat spectrum from the accretion shock model. 

The gamma-ray emission from accretion shocks is expected to have an irregular morphology since the cooling time of these relativistic electrons is much shorter than the dynamic time of the cluster \citep{Pfrommer2010}.  The brightest region in the source W may correspond to a  luminous peak where the accretion shock intersects the galaxy filament \citep{Keshet2003Shock}.
The kinetic energy flux through the surface of the accretion shock  is $\dot E_k=1/2 \rho V_{\rm sh}^3 S$ \citep{Brunetti2014IJMPD}, where $\rho=n m_p$  is the  the upstream gas density ($n$ is the number density of the upstream gas and $m_p$ is the proton mass),  $V_{\rm sh}$ is  the shock velocity  and $S$ is the shock surface area, respectively. Assuming that a
fraction $\eta_e$ of the kinetic energy flux through the shock-surface is transferred
to  relativistic cosmic-ray electrons, the total IC luminosity
from CR is    $L_{\rm IC}\simeq 1/2 \eta_e \rho V_{\rm sh}^3 S$.

The observed gamma-ray luminosity in 100 MeV-1 TeV is a fraction ${\rm ln(10^4)/ln(\gamma_{max}/\gamma_{min})}$ of the total IC luminosity for a flat spectrum of $\nu F_\nu \propto \nu^0$. Taking  $S=\pi R^2=2.8\times10^{49}{\rm cm^2}$ (assuming $R\sim 1{\rm Mpc}$ for the size of the shock surface of the source W) and ${\rm ln(\gamma_{max}/\gamma_{min})=20}$, we obtain 
\begin{equation}
    \eta_e \simeq 10^{-3} 
    \left(\frac{n}{10^{-4}{\rm cm^{-3}}}\right)^{-1}
    \left(\frac{V_{\rm sh}}{10^3 {\rm km~ s^{-1}}}\right)^{-3} \left(\frac{R}{\rm 1Mpc}\right)^{-2}.
\end{equation}
This efficiency is consistent with the numerical simulation for accretion shocks \citep{Ha2023}. It also agrees with the inferred acceleration efficiency from the giant radio relics in some galaxy clusters \citep{Brunetti2014IJMPD}.

\section{Conclusions and Discussions}

Structure formation in the intergalactic medium  produces large-scale, collisionless shock waves, in
which electrons can be accelerated to highly relativistic energies. These electrons can produce gamma-ray emission through IC scatterings of the CMB photons. We reported the detection of such gamma-ray emission at the outer region of the Coma cluster. The hard spectrum   of the excess emission, as well as its location in a large-scale filament of galaxies, are  consistent with the accretion shock origin for the gamma-ray emission.  

The IC emission could extend down to X-ray band as long as the power-law distribution of CR electrons extends to a minimum Lorentz factor of $\gamma_{\rm min}\sim 10^3$. Assuming $\alpha=-1.81\pm0.21$  for the photon index, we expected a differential flux of $F_{\rm X} \sim 10^{-14}-10^{-13}{~\rm erg ~cm^{-2} ~s^{-1}}$ at X-ray band for the source W. Deep observations with {NuSTAR}, eROSITA and Einstein Probe could test this prediction. 

The synchrotron radio emission of the source W can be estimated by
$L_{\rm radio} \sim L_\gamma (\frac{B}{B_{\rm CMB}})^2$, where $B_{\rm CMB}\equiv(8\pi T_{\rm CMB}^4)^{1/2}=3.2(1+z)\mu {\rm G}$ is defined as the magnetic field for which the magnetic energy density equals the CMB energy density. Assuming $B\le 0.1\mu {\rm G}$, we obtain $L_{\rm radio} \le 10^{-3} (B/0.1\mu {\rm G})^2 L_\gamma \sim 10^{39}{\rm erg~s^{-1}}(B/0.1\mu {\rm G})^2$. The radio emission of the source W could be also tested by sensitive radio observations in the future.

\begin{acknowledgments}
The work is supported by the NSFC under grants Nos. 12333006, 12121003 and 12203022.
\end{acknowledgments}

\appendix
\section{Systematic uncertainties}

To assess the robustness of our results, we investigate how the detection significance of  the source W changes in the systematic checks. 
We first study the impact caused by different low-energy thresholds of 100, 300, and 500 MeV. The spectral index and the detection significance of the source W change by less than 10\%, primarily due to the hard spectral index of the source W.

Although Coma is located near the galactic north pole (i.e., in a very clean region regarding  Galactic diffuse gamma-ray emission), the uncertainty from Galactic diffuse foreground modeling may still be significant.  We thus compare results obtained by using the standard diffuse emission model with those
obtained by using alternative diffuse emission models. \cite{FermiDiffuse2012} provides 128 sets of
maps corresponding to different model parameters. Following \citet{Xi2018Coma}, we
adopt 16 sets among them, which varies in the most
important parameters involved in creating the template,
including CR source distribution (Lorimer, SNR), halo size
(4, 10 kpc), spin temperature (150 K, 10$^5$ K), and E(B-V)
magnitude cut  (2, 5 mag). 
%\textbf{First, we find that our results do not change if we use the previous background model  {\tt\string gll\_iem\_v06}. We also consider  16 alternative background models as discussed in \cite{Xi2018Coma}. They are obtained using the GALPROP web interface for computation \citep{GALPROPWebRun}.  }
For isotropic spectral templates, we used 3 isotropic spectral templates and the corresponding instrument response function, {\tt\string iso\_P8R3\_SOURCE},  {\tt\string iso\_P8R3\_CLEAN} and {\tt\string iso\_P8R3\_UNTRACLEANVETO}  along with their corresponding the event class and event type.
The other two have larger effective areas at the expense of higher contamination of background events.
By testing different combinations of these templates, we found that the error introduced by background selection was below 25\%, with the detection significance (the TS value) ranging from 23.52 to 28.39 for the source W and from 45.72 to 62.25 for the disk component.

We summarize the results on systematic uncertainties in Table \ref{tab:uncertainties}. The largest uncertainty in the photon flux arises from the uncertainty in the Galactic diffuse foreground, while other effects have only minor influence.

\begin{table}
    \centering
    \begin{threeparttable}
    \begin{tabular}{l lccc}
    \hline
     Type  & Variation  & \  Spectral index impact \ & \ Flux impact \ & \ Range of TS$_{\rm Source W}$ \ \\
    \hline
     Diffuse modeling  & Alt. diffuse models\tnote{a}  & $< 7\%$ & $< 15\%$ 
&  (24.55, 28.39)\\
     Isotropic spectral template & Alt. isotropic models \tnote{b}  & $< 9\%$ & $< 20\%$   &  (23.52, 28.39)   \\
     Low-energy threshold & 100–500 MeV  & $< 8\%$ & $< 10 \%$ & (23.53, 25.66) \\
    \hline

    \end{tabular}
    \begin{tablenotes}
    \item[a] {\tt\string gll\_iem\_v07}, {\tt\string gll\_iem\_v06} and 16 alternative background models discussed in \cite{Xi2018Coma}.
    \item[b] {\tt\string iso\_P8R3} for three event classes {\tt\string SOURCE} (evclass = 128), {\tt\string CLEAN} (evclass = 256) and {\tt\string UNTRACLEANVETO} (evclass = 1024).
    \end{tablenotes}
    \end{threeparttable}
    \caption{Systematic uncertainties for the source W.}
    \label{tab:uncertainties}
\end{table}

\section  {Search for radio sources in the error box of the source W }  \label{appendix:radioS}
We study whether there is any counterpart to the source W. Since most of the gamma-ray-emitting AGNs are sources of radio emission, we search for possible radio sources  within 0.12$^\circ$ (99\% containment
radius) around the point-like source in the radio source catalogs of SIMBAD Astronomical Database \footnote{\url{http://simbad.cds.unistra.fr/simbad}} and NASA/ IPAC EXTRAGALACTIC DATABASE (NED) \footnote{\url{https://ned.ipac.caltech.edu/forms/nearposn.html}}.  
There are no AGNs  within 0.12$^\circ$. 
A bright radio AGN (7C 124442.50+275745.00) with a flux density of 178 mJy is located beyond the 99\% containment
radius given by the \textit{source\_localization} tool implemented in \textit{Fermipy}, so we think this source is irrelevant to the source W. 
We also check the flat-spectrum radio source catalog, CRATES \footnote{\url{https://heasarc.gsfc.nasa.gov/W3Browse/radio-catalog/crates.html}}. No source is found 
within our 99\% containment radius.

% Supplemented in Appendix B
In our search for possible X-ray counterparts in the eROSITA DR1 catalog \footnote{\url{https://erosita.mpe.mpg.de/dr1/erodat/}}, we found a  X-ray point source, 1eRASS J124657.2+272806, within the $1\sigma$ positional error of the source W, with a flux of $1.2 \times 10^{-13} \, {\rm erg \, cm^{-2} \, s^{-1}}$. Additionally, by comparing this region with data from the VLASS survey \footnote{\url{https://science.nrao.edu/vlass}}, we note the presence of a radio point source, VLASS1QLCIR J124657.34+272759.6, within the $2\sigma$ positional uncertainty region of the eROSITA X-ray source, with a flux density of \(1.4 \, {\rm mJy}\) in the 2–4 GHz band. According to the radio-GeV flux correlation for {\em Fermi} blazars \citep{Ghirlanda2011}, the expected GeV flux of this radio source is insufficient to account for the GeV flux of source W, although we cannot completely rule out  the possibility of the association between two sources considering the uncertainty in the radio-GeV correlation.

% Four faint unidentified radio sources in the SIMBAD Astronomical Database are located within the 99\% containment radius. The radio flux densities of three of them  are 
% 3.5 $\pm$ 0.6 mJy for NVSS J124653+272912, 
% 4.4 $\pm$ 0.5 mJy for NVSS J124708+273102 and
% $\sim$ 8.32 mJy for FIRST J124732.3+273111 at 1.4 GHz. We can estimate the gamma-ray luminosity of these sources if the gamma-ray emission is mostly due to Compton scattering of the radio-producing electrons by the CMB, a reasonable expectation in light of the conclusion reached by Ref. \cite{Fermi2010Sci} in their analysis of Fermi-LAT measurements of the radio galaxy Cen A. We find that the expected gamma-ray flux too low to explain the observed flux  of the source W.
% Another radio source, NVSS J124706+272800, has a radio flux density of 37.9 mJy at 1.4GHz and 0.3 Jy at 74 MHz respectively. Such a soft spectrum suggests that it is highly unlikely to be a blazar \cite{Planck2011}.

\section{Flux variability of the source W}  \label{appendix:FluxVar}

\begin{figure}
    \centering
    \includegraphics[width=0.8\linewidth]{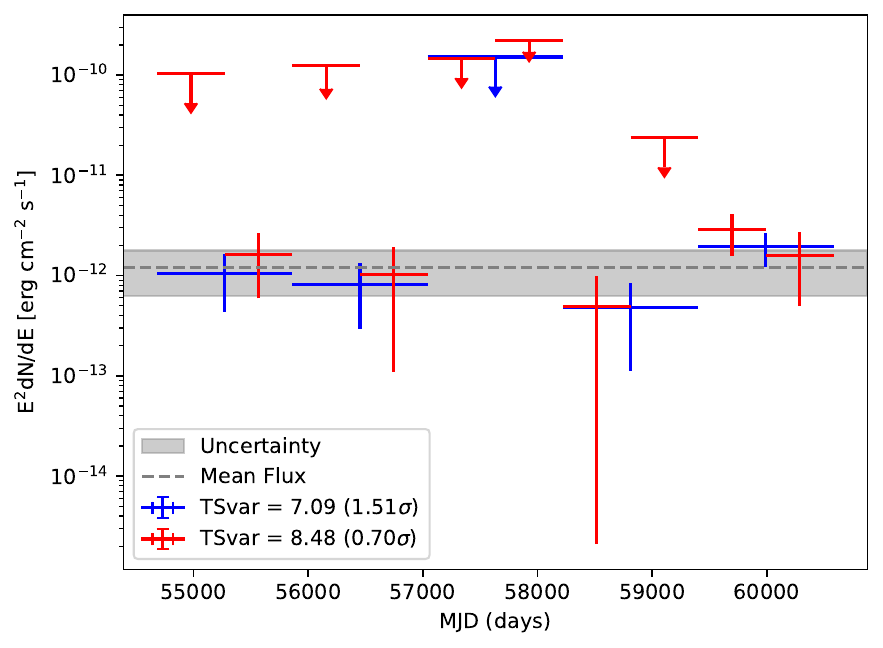}
    \caption{Light curves of the source W with five and ten time bins. The mean flux is the averaged flux over  16.2 yr. The upper limits at the 95\% confidence level are derived when the TS value for the data points is lower than 4.}
    \label{fig:LC}
\end{figure}

We retained the point-source model at the best-fit location for
examining the variability of the gamma-ray flux. We computed light
curves in five and ten time bins using 16.2 years of Fermi-LAT data, for events in the
energy range 0.1-1000 GeV. For the analysis in each time bin, all
sources within the 6$^\circ$ region around the source W have their spectra
fixed to the shapes obtained from the above broadband analysis.
The result is shown in Fig. \ref{fig:LC}. We then used a likelihood-based
statistic to test the significance of the variability. Following the
definition in 2FGL \cite{Nolan2012ApJS}, the variability index from
the likelihood analysis is constructed, with a value in the null
hypothesis where the source flux is constant across the full time period, and the value under the alternate hypothesis where the
flux  in each bin is optimized: ${\rm TS}_{\rm var} =2\times \sum_{i=1}^N [ \log(\mathcal{L}_i(F_i)) -\log(\mathcal{L}_i(F_{\rm mean})) ] $  ($\mathcal{L}_i$ is the likelihood corresponding to bin
\textit{i}, $F_i$ is the best-fit flux for bin \textit{i}, and $F_{\rm mean}$ is the best-fit flux for
the full time) \cite{Nolan2012ApJS}. We find a $ 0.7 - 1.5 \sigma$
significance for the flux variability for the analyses using the
above two time bins, which suggests no significant variability for
the gamma-ray emission from the source W.

\bibliography{sample631}{}
\bibliographystyle{aasjournal}

\end{document}